\documentclass{elsart}

\begin{document}
\baselineskip1.0cm
\begin{frontmatter}
\title{Determination of Scanning Efficiencies in Experiments Using Nuclear
Emulsion Sheets}
\author[gb]{G.~Brooijmans}
\address[gb]{Universit\'{e} Catholique de Louvain, Louvain-La-Neuve, Belgium.\\
        Institut Interuniversitaire des Sciences Nucl\'eaires, Belgium.
        \thanksref{gbn}} 

\thanks[gbn]{Now at Fermi National Accelerator Laboratory, Batavia, IL.}

\begin{abstract}
During their exposure, nuclear emulsion sheets detect both tracks from
experiment-related particles, as well as a considerable amount of background
tracks, mainly due to cosmic rays.  Unless the exposure has been fairly short, 
it is therefore
fairly likely that a fraction of the tracks that have been identified as 
belonging to the particles the experiment is interested in, are really due to 
background.  A method, which allows to measure this fraction 
reliably directly from the data, is described.
\end{abstract}
\end{frontmatter}

\newpage

\section{Introduction}
\label{intro}

Nuclear emulsion sheets are composed of two thin layers of nuclear emulsion
deposited on both sides of a transparent plastic base.  They offer track position
resolution of about 1 $\mu m$ and angular resolution around 1 milliradian.

Various present \cite{chorus,donut} and future neutrino \cite{tosca,opera} 
experiments have used, or plan to use, the excellent resolution offered
by nuclear emulsion sheets to unambiguously detect the production of a tau lepton.

These experiments typically use an electronic detector for calorimetry, 
particle identification and moderate precision tracking.  The tracking 
detectors are used to reconstruct the points where particles left
the emulsions.  Tracks compatible with the ``prediction'' are then 
searched for with microscopes, in an area proportional in size to the 
impact point prediction uncertainty.  If found, the tracks can be followed
from emulsion sheet to emulsion sheet until the interaction vertex is located,
or they are lost.

During exposure, 
which typically lasts for months, these emulsion sheets accumulate 
significant quantities of background, mainly due to cosmic muons.  While
most of this background is perpendicular to the beam direction, a small
fraction can be mistakenly identified as corresponding to one of the 
particles the experiment has reconstructed.  This generally leads to a
wrong estimation of the track finding efficiency in the emulsion sheets.
A method has been developed to determine the real track finding efficiency
from the effective observed efficiency directly from the available data.
The method does not involve any simulation.

In section \ref{sec:definitions} the real and effective track finding 
efficiencies are defined.  Then, in section \ref{sec:assume}, a few
assumptions are made which are usually true in neutrino experiments.  The
method to extract the real track finding efficiency is described in
section \ref{sec:method}, followed by a summary of the possible sources 
of uncertainties in section \ref{sec:uncert}.  Conclusions are drawn in
section \ref{sec:conclude}.

\section{Definition of Real and Effective Track Finding Efficiencies}
\label{sec:definitions}

The real track finding efficiency is defined by
\begin{equation}
\epsilon_{r} = \frac{Number \ of \ times \ the \ right \ track \ is \ found}
                    {Number \ of \ tracks \ searched \ for},
\end{equation}
and the observed track finding efficiency by
\begin{equation}
\epsilon_{o} = \frac{Number \ of \ times \ a \ matching \ track \ is \ found}
                    {Number \ of \ tracks \ searched \ for}.
\end{equation}

It is clear that at first sight the real track finding efficiency is unknown:  
the observed efficiency results from a convolution of signal and background, 
so that an unfolding procedure seems to be necessary to extract the real efficiency.  

\section{Assumptions}
\label{sec:assume}

The following assumptions are made:
\begin{itemize}
\item The random background can be assumed to be evenly distributed for all the 
tracks searched for in the emulsion sheets.  This assumption holds for emulsion
sheets of uniform quality which have been exposed during the same period.
\item The real track finding  efficiency can have only two results: found or not found.
This is true if the density of events is low enough, and if particles from one event 
are sufficiently separated in phase space.  This condition is satisfied in the 
experiments considered here.
\item For each track, an area of identical size (proportional to the 
prediction accuracy) is scanned.  
In other words, when a track
candidate is found, the rest of the area must still be scanned for possible additional
candidates.
\item The maximal number of candidates for a single track exit point prediction 
is finite.  This is an assumption on the emulsion quality and the experiment's 
ability to see tracks in the emulsion.
\end{itemize}

An important consequence of the first and third assumptions is that the number of 
tracks that migrate from 0 candidates to $n$ candidates
due to the background is equal (within statistics) to the number of tracks that migrate 
from 1 to $n+1$ candidates.
This relies on the even distribution of background and the scanning of the entire area.

\section{The Method}
\label{sec:method}

The idea is simple: the real result can be only zero (not found) or one (found), while
in the data the number of good candidates in a given surface can be much larger due to
background.


Consider the distribution of the number of candidate tracks found for tracks
that were searched for in an emulsion sheet.  In a healthy experiment, this distribution
has a few entries at zero, a large peak at 1, and has a small tail extending to
larger values.

Call $N'(i)$ the number of entries in bin $i$, corresponding to the number of tracks 
for which $i$ candidates were found.  And call $N(i)$ the number of entries 
that would be observed in each bin for the real track finding efficiency.  
Under the second assumption $N(i) = 0$ for $i > 1$.

Then let $X(i \rightarrow j)$ be the number of entries that have migrated from one
bin to another due to the background.  Here $i$ is necessarily 0 or 1 due to the second
assumption, $j > 0$ and $j > i$.

Since $X(0 \rightarrow j) = 0$ for some arbitrarily large $j$ (last assumption),
\begin{eqnarray}
X(1 \rightarrow j) = X(0 \rightarrow (j-1)) & = & N'(j)\ , \nonumber \\
X(1 \rightarrow (j-1)) = X(0 \rightarrow (j-2)) & = & N'(j-1) - N'(j)\ , \nonumber \\
\ldots \ and \nonumber \\
X(1 \rightarrow 2) = X(0 \rightarrow 1) & = & N'(2) - N'(3) + N'(4) - \ldots \label{eq:xval}
\end{eqnarray}
But, by construction,
\begin{equation}
N(0) = N'(0) + \sum_{k} X(0 \rightarrow k) \ , 
\end{equation}
which, using (\ref{eq:xval}), is equivalent to
\begin{equation}
N(0) = \sum_{l=0}^{j/2} N'(2 l) \ .
\end{equation}

Therefore the real track finding efficiency is
\begin{equation}
\epsilon_{r} = \frac{N_{tot} - N(0)}{N_{tot}} = 
\frac{N_{tot} - \sum_{l=0}^{j/2} N'(2 l)}{N_{tot}},
\end{equation}
where $N_{tot}$ is the total number of tracks searched for.

\section{Uncertainties}
\label{sec:uncert}

The statistical uncertainty on the contents of each bin quantifies the error on 
\begin{equation}
X(1 \rightarrow n) = X(0 \rightarrow (n-1)),
\end{equation}
leading to a statistical uncertainty equal to the quadratic sum of the 
individual statistical uncertainties of the even bins.  

Systematic uncertainties can be induced by the scanning method and individual
emulsion sheet properties, so that it is recommended to treat sheets one by one.

\section{Conclusion}
\label{sec:conclude}

A method to extract the real track finding efficiency from the observed track 
finding efficiency in emulsion sheets has been described in detail.  It relies
on assumptions which are not stringent and usually satisfied in any experiment
using these types of sheets.

\ack

The author acknowledges the financial support of the Institut 
Interuniversitaire des Sciences Nucl\'eaires (Belgium).

\end{document}